\documentclass[manuscript]{aastex}

\usepackage{graphicx}   
\usepackage{amsmath}    
\usepackage{xcolor}     

\shorttitle{Spherical ChromaStar}
\shortauthors{Short}

\begin{document}


\title{Chroma+ model stellar surface intensities: Spherical formal solution}


\author{C. Ian Short}
\affil{Department of Astronomy \& Physics and Institute for Computational Astrophysics, Saint Mary's University,
    Halifax, NS, Canada, B3H 3C3}
\email{ian.short@smu.ca}





\begin{abstract}

We announce V. 2025-08-08 of the Chroma+ suite of stellar atmosphere and spectrum
modelling codes for fast, approximate, effectively platform-independent stellar 
spectrum synthesis, written in a number of free well-supported  programming languages. 
The Chroma+ suite now computes the emergent surface intensity and flux distributions
and the hydrostatic pressure structure assuming a spherical atmosphere rather than
local flatness by implementing the analytic formal solution of the $1D$ spherical
radiative transfer equation of \citet{chapman66} based on an integrating factor.  
We present our adaptation and discretization of the solution and demonstrate the 
resulting impact of our sphericity treatment on a number of computed observables,
including exo-planet transit light-curves.
All codes are available from the OpenStars www site:
www.ap.smu.ca/OpenStars.
 
\end{abstract}



\section{Introduction}

\paragraph{}

The Chroma+ stellar atmosphere and spectrum modelling suite (Chroma+, \citet{shortb21} and papers in that series)
provides for quick self-consistent convergence of the total hydrostatic pressure structure, $P(\tau_{\rm Ros})$, 
the gas $P_{\rm gas}(\tau_{\rm Ros})$, electron $P_{\rm e}(\tau_{\rm Ros})$, and radiation $P_{\rm rad}(\tau_{\rm Ros})$ pressure 
structures, the opacity
structure, $\kappa_\nu(\tau_{\rm Ros})$ and $\kappa_{\rm Ros}(\tau_{\rm Ros})$, the gas pressure equation-of-state, 
$P_{\rm gas}(T_{\rm kin}, \rho, P_{\rm e}, ...)$ (EOS), and
the molecular and ionization equilibria for a given kinetic temperature structure, $T_{\rm kin}(\tau_{\rm Ros})$,
and synthesizes the spectrum with the VALD (\citet{VALD19}) atomic line list and select molecular bands in the just-overlapping-line
approximation (JOLA, \citet{jola}).
The  $T_{\rm kin}(\tau_{\rm Ros})$ structure, along with inital guesses at the  $P_{\rm gas}(\tau_{\rm Ros})$
and  $P_{\rm e}(\tau_{\rm Ros})$ structures, is scaled from a suitable starting model properly converged with 
V. 15 of the Phoenix (\citet{allard95}) atmospheric 
modeling code, thus allowing us to avoid the need to converge the line-blanketed radiative thermal equilibrium 
solution.  The Chroma+ suite has been provided in effectively platform-independent languages including Python
(ChromaStarPy {DOI: zenodo.1095687}), Java, 
and Javascript, making it suitable for quick, easily accessible numerical experiments in a pedagogical context, 
as well as for research contexts where parameter-perturbation and differential comparison are central.
The Chroma+ suite may be compared to the Spectroscopy Made Easy (SME) package described in \citet{SME1} and \citet{SME2}
under IDL.  
The suite is available from the OpenStars www site: www.ap.smu.ca/OpenStars.

\paragraph{}

The Chroma+ suite originally adopted the simplifying approximation of local flatness, or plane-parallel (PP) geometry, in which the 
atmosphere is treated as a planar slab of gas infinite in extent perpendicular to the line-of-sight, normally taken to be 
the $z$-axis of the Cartesian and spherical polar coordinate systems.
The PP approximation is 
most realistic for stars of relatively large surface gravity, $\log g$, and becomes less realistic with decreasing $\log g$ value.  
In modelling the outgoing surface specific intensity distribution,
$I^+_\lambda(\tau=0, \cos\theta)$, where $\theta$ is the angle between the  $I^+_\lambda$ beam and the surface normal 
of a PP model, the approximation is best for the 
pencil beam normal to the star's surface ($\cos\theta = 1$) and worsens as beams approach 
grazing-incidence ($\cos\theta \gtrsim 0$).  For a spherical star, beams of sufficiently small $\cos\theta$ value pass through a 
geometric path length, $\Delta z$, that corresponds to an optical depth interval, $\Delta\tau$, less than unity so that 
the star is optically thin for that beam, whereas in PP geometry all beams emerge from a semi-infinite 
medium. The effects of the PP approximation on various observables and on $1D$ vertical atmospheric structure have been
well documented (\citet{flr90}, \citet{neilson13a}).  One effect of more recent concern is that on the
modelled light-curves of exo-planetary transits (\citet{neilson17}, \citet{neilson22}), particularly during ingress and egress during which
$I^+_\lambda(\tau=0, \cos\theta)$ beams of $\cos\theta \gtrsim 0$ are being occulted. 

\paragraph{}

Because the Chroma+ suite does not converge the line blanketed radiative thermal equilibrium problem,
we only need account for sphericity in the formal solution of the radiative transfer equation (RTE) for the 
 $I^+_\lambda(\tau=0, \mu_0)$ distribution for a given $T_{\rm kin}(\tau)$ structure,
where $\mu_0\equiv\cos\theta$ for $\theta$ defined at the surface.
As of Version 2025-08-08 (ISO 8601 versioning), the Chroma+ suite now evaluates the formal solution for the $I^+_\lambda(\tau=0, \mu_0)$ 
distribution
adopting spherical atmospheric geometry by discretizing the analytic formal solution of \citet{chapman66} based on an
integrating factor.
We note that our
$T_{\rm kin}(\tau)$ structures are re-scaled from Phoenix V. 15 \citet{allard95} models that were calculated with spherical geometry, 
so the adoption of sphericity in our formal solution makes our overall procedure more self-consistent.
Sphericity also affects the formal solution of the hydrostatic equilibrium equation (HSE) for the total
pressure structure, $P(\tau)$ (\citet{flr90}), 
albeit in a way that is relatively straightforward
to accommodate.   For consistency, we also now allow for sphericity in our HSE solution.

\paragraph{}
There are other codes, written in FORTRAN, that solve the spherical radiative transfer problem, such as S3R2T (V15), the radiative transfer module of Version 15 of
the PHOENIX stellar atmosphere and spectrum modelling code
(\citet{allard95}) and SATLAS (\citet{lestern08}), the spherical version of the ATLAS9 and ATLAS12 atmospheric
modelling codes (\cite{kurucz14}, \citet{castellik06}).
 However, the Chroma+ suite has the advantage of being
very fast because of its approximate approach for economizing the procedure, and is suitable for more interactive environments such
as the Python integrated development environment (IDE), which allow a user to more quickly extract approximate results from
fitting observed spectra, and the Javascript/HTML environment which allows for web demonstrations.  By contrast,
the spectrum synthesis procedure in the Java version of the Chroma+ suite has been parallelized in the 
wavelength domain and allows for relatively fast spectrum synthesis of large spectral regions blanketed by thousands of lines.
All versions have been updated with the spherical solution.

\paragraph{}

In Section \ref{basics}, we review the radiative transfer formal solution of \citet{chapman66} and our  
adaptation and discretization of the solution. 
In Section \ref{results} we present the comparison of the emergent observables computed 
with spherical and PP geometry.

\section{The spherical formal solution}
\label{basics}

\subsection{Radiative transfer equation (RTE)}

The monochromatic RTE for the monochromatic specific intensity distribution, $I_\lambda(r, \mu)$, in $1D$ spherical geometry 
must be formulated with absolute radial height, $r$, as the independent position coordinate and is

\begin{equation}
   {{\mu}\over{\kappa_\lambda(r)\rho(r)}}{{dI_\lambda(r, \mu)}\over {dr}} + {{(1-\mu^2)}\over {\kappa_\lambda(r)\rho(r) r}}{{dI_\lambda(r, \mu)}\over {d\mu}} = S_\lambda(r) - I_\lambda(r, \mu)   
\end{equation}

 where $\mu$ is now a variable that depends on $r$ for a given $I_\lambda$ beam and is the cosine of the angle between the beam
 and the normal to the tangent to the local concentric spherical shell being intersected by the beam, and $S_\lambda$ is the radiative 
 source function and is assumed to be isotropic ($S_\lambda(\mu) = S_\lambda$).

 \paragraph{}

\citet{chapman66} (C66 henceforth) presents an analytic expression for the formal solution for an atmosphere of infinite extent that
was derived using an integrating factor, $\Phi(r, \mu)$.  Below we present our adaptation of the solution
for a discretized atmospheric model of finite extent for outgoing surface intensity only, $I^+_\lambda(\tau=0, \mu_0)$.

\paragraph{}

In what follows, indices $i$ and $j$ refer to absolute radial height, $r$ ($\xi$ in the notation of C66),
$N$ is the total number of discrete height points and is currently set equal to 64,
and index $k$ refers to the $\mu_{0, k} \equiv \cos\theta_k$ value between the
direction of the current pencil beam ({\it ie.} the direction of integrating) and the radial direction at the {\it surface}
of the model (index $i=N$).  
The set $\{\mu_{0, k}\}$ is the 32 positive abscissae values of a 64-point Gauss-Legendre quadrature on the domain $[0, 1]$. 

\paragraph{}

We define the core to be a sphere of radius equal to the nominal stellar radius corresponding to the input
$\log g$ and $M$ parameters, so that the core (or ''inner'') radius, $R_i$ in the notation of C66, is $(GM/g)^{1/2}$.
Our independent depth variable is the radial Rosseland optical depth scale, $\tau_{\rm Ros}$, equally spaced in $\log\tau_{\rm Ros}$
in the range $[-6, 2]$.  Our established procedure computes the Rosseland mass extinction coefficient, $\kappa_{\rm Ros}(\tau_{\rm Ros})$,
from the total continuous extinction and then computes the resulting geometric depth scale, $z(\tau_{\rm Ros})$. 
To avoid unphysical values arising from the {\it ad hoc} upper boundary condition, we take the atmospheric radial extent, $\Delta z$, to
be $(z_0 - z_{N-2})$.  We then approximate the radius of the outermost layer of the atmosphere,
$R_0$ in the notation of C66, as $R_i + \Delta z$.  We then generate a grid of absolute discrete radial heights in our atmospheric model,
$\{ r_i\}$, from our $\{z_i \}$ values in the $i$ range $[2, N]$ by setting $r_0$ equal to $R_i$. 

\paragraph{}

In what follows, $r_b$ is the impact parameter of a parallel pencil beam travelling toward the observer and $b$ is the height index of the 
shell grazed by a beam, $k$.  We evaluated $r_b$ as

\begin{equation}
   r_b = r_N(1-\mu^2_{0, k})^{1/2}
\end{equation}
 
Following C66, we must distinguish between core-intersecting beams ($r_b < r_0$) and non-core-intersecting beams ($r_b \ge r_0$), and $\mu_c$   
is the critical value of $\mu_{0, k}$ distinguishing the two regimes in the notation of C66.  For the special case of
surface intensity ($r_i = r_N$), the critical value is given by 

\begin{equation}
	\mu_c = | [ 1 - ({r_0\over r_N})^2 ]^{1/2} |
\end{equation}

We replace the combination $\kappa_\lambda(\xi)\rho(\xi)d\xi$ in the integrating factor and the formal solution of C66 with our 
corresponding monochromatic $\Delta\tau_{\lambda, i}$ interval because our $\tau_\lambda$ scales are physically consistent with the
model structure and our computed $\kappa_\lambda(r)$ distribution. 
Then, the integrating factor for height $i$ and any local direction cosine, $\mu$, comparable to Eq. 25 of C66, is evaluated as 

\begin{equation}
	\Phi_\lambda(r_i, \mu) \approx \exp\{ -\sum_{j=i}^N {{\Delta\tau_{\lambda, \rm j}} \over {[ 1 - ({r_i\over r_ j})^2(1-\mu^2)]}^{1/2}} \}
\label{integFac}
\end{equation}

For the special case of the surface, $\Phi_\lambda(r_N, \mu) = \exp(-0) = 1$, simplifying the solution below.

\paragraph{}

We assume local thermodynamic equilibrium (LTE) so that the monochromatic source function, $S_\lambda(r_i)$, is given by the
Planck function at the local kinetic temperature, $B_\lambda(T_{\rm kin}(r_i))$.  
The surface intensity for out-going beams that are core-intersecting ($\mu_c \le \mu_{0, k} \le 1$), comparable to Eq. 28 of C66, is evaluated as

\begin{multline}
  I_\lambda^+(r_N, \mu_{0, k}) \approx B_\lambda(r_0)\Phi_\lambda\{r_0, {[ 1 - ({r_N\over r_0})^2(1-\mu_{0, k}^2)]}^{1/2} \}\\
	+ \sum_{i=1}^N \Phi_\lambda \{r_i, {[ 1 - ({r_N\over r_ i})^2(1-\mu_{0, k}^2)]}^{1/2} \}B_\lambda(r_i){{\Delta\tau_{\lambda, \rm i}}\over {[ 1 - ({r_N\over r_i})^2(1-\mu_{0, k}^2)]}^{1/2}} 
\label{outCoreI}
\end{multline}

The surface intensity for out-going beams that are non-core-intersecting ($0 \le \mu_{0, k} < \mu_c$), comparable to Eq. 29 of C66, is evaluated as

\begin{multline}
  I_\lambda^+(r_N, \mu_{0, k}) \approx I_\lambda^-(r_b, 0) \Phi_\lambda\{r_b, 0 \}\\
	+ \sum_{i=b}^N \Phi_\lambda\{r_i, {[ 1 - ({r_N\over r_i})^2(1-\mu_{0, k}^2)]}^{1/2} \}B_\lambda(r_i){{\Delta\tau_{\lambda, \rm i}}\over {[ 1 - ({r_N\over r_i})^2(1-\mu_{0, k}^2)]}^{1/2}}
\label{outNonCoreI}
\end{multline}

where $I_\lambda^-(r_b, 0)$ on the RHS of Eq. \ref{outNonCoreI}, comparable to Eq. 27 of C66, is evaluated as

\begin{equation}
	I_\lambda^-(r_b, 0) \approx \sum_{i=b}^N ({{\Phi_\lambda\{r_b, 0\}}\over {\Phi_\lambda\{r_i, [ 1 - ({r_b\over r_i})^2]^{1/2}\}}})\\
	B_\lambda(r_i){{\Delta\tau_{\lambda, \rm i}}\over {[ 1 - ({r_b\over r_i})^2(1-\mu_{0, k}^2)]}^{1/2}}
\label{inNonCoreI}
\end{equation}

\subsection{Hydrostatic equilibrium (HSE)}
\label{HSE}

We also account for the effect of sphericity on the hydrostatic pressure structure, $P(r)$.  Following \citet{lestern08} and \citet{flr90} the HSE is now 

\begin{equation}
	{{dP}\over{d\tau_{\rm Ros}}} = { {g(\tau_{\rm Ros})}\over {\kappa(\tau_{\rm Ros})} }
   \label{HSEEq}
\end{equation}

 where $g(r)$ is evaluated as $GM/r^2$ with the stellar mass, $M$, held constant at its total value and the $\{ r_i\}$ grid is computed
 as described above.

 \paragraph{}

 We adapt the formal solution of \citet{gray3} that we numerically integrate inward so that it now accounts in the integrand
 for the dependence of the gravitational acceleration, $g$, on the radial height, $r$,

 \begin{equation}
	 P(\tau_{\rm Ros, i}) = \{ {3\over2}\sum_{j=N}^i  {{\tau_{\rm Ros, j} g(\tau_{\rm Ros, j})P_j^{1/2}} \over {{\kappa_{\rm Ros, j}}}} \Delta\log\tau_{\rm Ros, j}\}^{2/3} 
 \label{HSESoln} 
 \end{equation}

  where the $\{P_i\}$ values on the RHS are the current guess of the total pressure structure.  

\subsection{Modeling parameters}

 We now need an additional independent input parameter in our modelling procedure to evaluate $r_0$, and we have chosen stellar mass, $M$.  
 The $r_0$ value is then obtained from the input value of $\log g$ as $\log r_0 = 0.5(\log G + \log M - \log g)$.  We have provided a new
 Boolean input parameter, \texttt{ifSphere}, that allows the user to select between the spherical and PP 
 formal solutions of the RTE and the HSE.

\section{Results}
\label{results}

\begin{figure}
\includegraphics[width=\columnwidth]{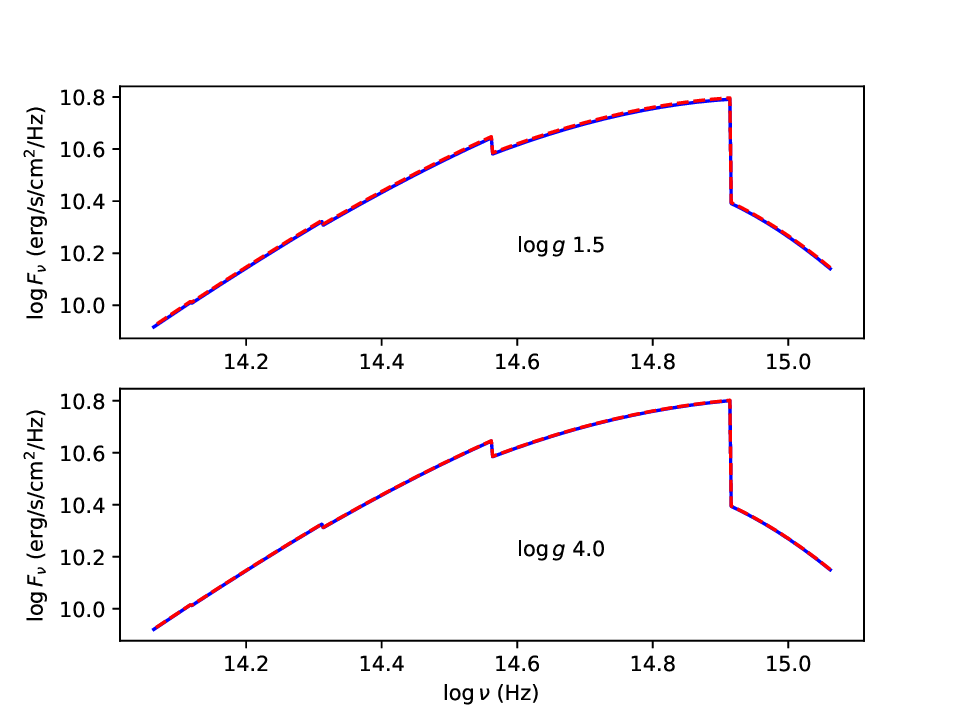}
	\caption{Spectral energy distribution (SED) for the test models of \citet{flr90} of
$T_{\rm eff}$ of $10\,000$ K, $M$ of 2.0 M$_{\rm Sun}$ and $\log g$ 4.0 (upper panel) 
and 1.5 (lower panel)
 computed with the spherical solution (blue solid line) and the PP 
 solution (red dashed line), comparable to Fig. 2 of  \citet{flr90}.
  \label{FLR90Sed}
}
\end{figure}

\citet{flr90} implemented a spherical formal solution for the HSE, RTE and the radiative thermal equilibrium solution, in the ATLAS
code (\citet{kurucz14}) and studied the
effect on the modeled structure and emergent spectral energy distribution (SED) of a set of models of an early-type star of
$T_{\rm eff}$ value equal to $10\, 000$ K with varying $\log g$  and $M$ values.  In Fig. \ref{FLR90Sed} we show our computed
SEDs ($\log F_\nu$ {\it vs} $\log \nu$) for models of $T_{\rm eff}$ value equal to $10\, 000$ K, $M$ equal to 2.0 $M_{\rm Sun}$ and 
$\log g$ values of 4.0 and 1.5, 
computed in PP and spherical geometry for comparison to Fig. 2 of \citet{flr90}.  The SEDs for the models of $\log g$
value of 4.0
are indistinguishable in this plot, which is to be expected at relatively high $\log g$ values.  For the  models of $\log g$
value of 1.5, the SEDs are only barely distinguishable in this plot, with the most visible difference being that the PP SED is
slightly greater than the spherical SED 
in the broad region where $F_\nu$ peaks.  We emphasize that our models all have the same
$T_{\rm kin}(\tau)$ structure and only differ in the solution to the formal solutions of the RTE and HSE, whereas \citet{flr90}
are comparing SEDs for atmospheric models in which the radiative equilibrium $T_{\rm kin}(\tau)$ structure is also converged
consistently with the geometry.

\begin{figure}
\includegraphics[width=\columnwidth]{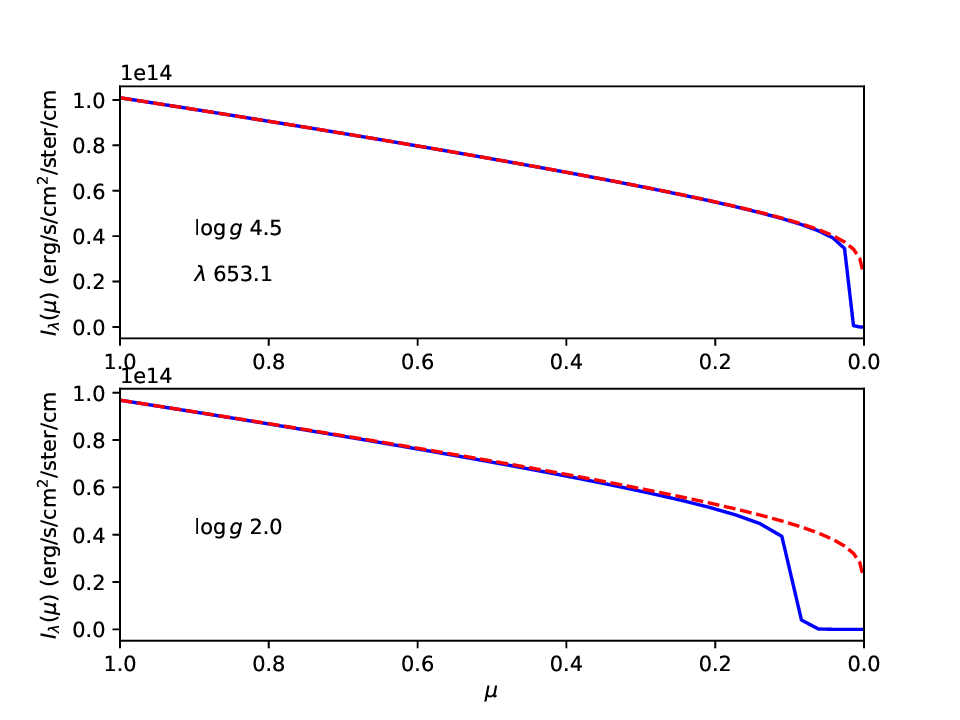}
\caption{The continuum limb darkening curve for the test model of \citet{neilson13a} of
 $T_{\rm eff}$ of 5000 K, $\log g$ of 2.0, and $M$ of 5.0  M$_{\rm Sun}$
 (lower panel)  computed with the spherical solution (blue solid line) and the PP 
 solution (red dashed line).  For comparison we also include the limb darkening 
 for a model of the same parameters except for a $\log g$ value of 4.5
 (upper panel).  The limb darkening is shown for a $\lambda$ value from
our background continuum $\lambda$ grid of 653.1 nm, near the centre of the Kepler bandpass,
 and is comparable to Fig. 1 of \citet{neilson13a}. 
  \label{CompareNeilsonLDC}
}
\end{figure}

\citet{neilson13a} compared the centre-to-limb intensity variation (CLIV) and the fitted parameters of 
a number of limb darkening laws for spherical and PP models based on the ATLAS code \citet{kurucz14}
in a number of photometric bandpasses, including that of the {\it Kepler} space telescope, and their
investigation includes a model of $T_{\rm eff}$ equal to 5000 K. $\log g$ equal to 2.0, and $M$ equal
to 5.0 M$_{\rm Sun}$.
In Fig. \ref{CompareNeilsonLDC} we show our CLIV computed with the Chroma+ suite for a model of these
parameters with the spherical and PP 
formal solutions at a wavelength from our background continuum $\lambda$-grid of 653.1 nm, close to the
centre of the {\it Kepler} bandpass.  This Figure may be compared with Fig. 1 of \citet{neilson13a}.
Our results are at least qualitatively similar to theirs in that we find the greatest
difference between the two geometries is near the limb where $\mu_0$ is less than $\sim 0.1$, with 
the spherical CLIV dropping precipitously below that from the PP model.   This is to be expected - at 
low $\mu_0$ values near the limb, the $I^+_\lambda(\tau=0, \mu_0)$ beams are emerging from an optically
thick path-length in a semi-infinite planar atmosphere, whereas for the spherical model they are 
emerging from a finite path-length through a spherical shell that does not intersect the opaque inner core. 

\paragraph{}

\begin{figure}
\includegraphics[width=\columnwidth]{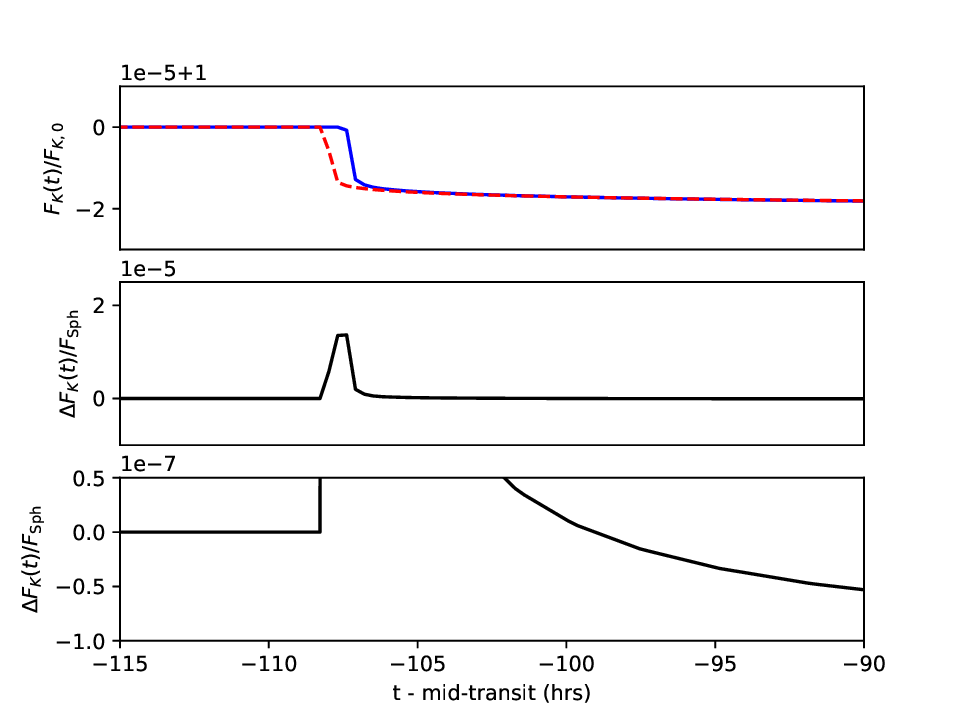}
\caption{Upper panel: The transit light-curve at ingress for the model of Fig. \ref{CompareNeilsonLDC} and
a planet of radius equal to $1 R_{\rm Jup}$ and orbital radius equal to 1 AU 
 computed with the spherical solution (blue solid line) and the PP
 solution (red dashed line).  
Middle and lower panels: The relative difference between the light-curve computed with the
spherical and the PP model.
  \label{CompareNeilsonLightCurve}
}
\end{figure}

To draw out the implications for predicted exo-planetary transit light-curves, in Fig. \ref{CompareNeilsonLightCurve}
we show the predicted transit light-curves for a host star of the same parameters and a planet of radius 1 R$_{\rm Jup}$, a circular orbit
of radius 1 AU, and an orbital inclination, $i$, of 90$^{\rm o}$.  For the spherical geometry, the apparent onset
of ingress is delayed, and the light variation at ingress is more rapid, than that for PP geometry, and this is
consistent with the more rapid decline in $I^+_\lambda(\tau=0, \mu_0)$ with decreasing $\mu_0$ at small
$\mu_0$ values near the limb in the spherical model as compared to the PP model.

\section{Discussion }
\label{discussion}

The Chroma+ suite may now be used for projects in which the student investigates the limb darkening of exo-planetary host 
stars with $I^+_\lambda(\tau=0, \mu_0)$ distributions computed with spherical and planar geometry and the corresponding 
exo-planetary transit light-curves.  This includes projects in which students investigate the realism of various 
limb-darkening laws, find best-fit limb darkening coefficients for various laws, and investigate the effect on 
inferred exo-planet and host star properties (\citet{neilson13a}, \citet{neilson13b}).

\end{document}